\documentclass[9pt,twocolumn,twoside]{osajnl}

\journal{josab} 

\setboolean{shortarticle}{false}

\usepackage{subcaption}
\usepackage{tikz}
\usepackage{mathtools}
\usepackage{dblfloatfix}

\usetikzlibrary{patterns,decorations.pathreplacing}
\usetikzlibrary{arrows.meta}
\usetikzlibrary{shapes.arrows, fadings}

\let\originalleft\left
\let\originalright\right
\renewcommand{\left}{\mathopen{}\mathclose\bgroup\originalleft}
\renewcommand{\right}{\aftergroup\egroup\originalright}

\providecommand{\df}{\textrm{d}} 
\newcommand{\diff}[3][\hspace{-0.5pt}]{\frac{\df^{#1}#2}{\df{#3}^{#1}}} 
\newcommand{\pdiff}[3][\hspace{-0.5pt}]{\frac{\partial^{#1}#2}{\partial{#3}^{#1}}} 
\newcommand{\Es}{E_{\textrm{sat}}} 
\newcommand{\FT}[1]{\mathcal{F}\left\{ #1 \right\}} 
\newcommand{\FTi}[1]{\mathcal{F}^{-1}\left\{ #1 \right\}} 

\providecommand{\bigO}[1]{\ensuremath{\mathop{}\mathopen{}\mathcal{O}\mathopen{}\left(#1\right)}} 

\DeclareMathOperator{\sech}{sech}

\title{Predicting instabilities of a tuneable ring laser with an iterative map model}

\author[1,3]{Brady Metherall}
\author[2,4]{C. Sean Bohun}

\affil[1]{Mathematical Institute, University of Oxford, Radcliffe Observatory Quarter, Woodstock Rd, Oxford OX2 6GG, UK}
\affil[2]{Faculty of Science, University of Ontario Institute of Technology, 2000 Simcoe St N, Oshawa, ON L1G 0C5, Canada}

\affil[3]{brady.metherall@maths.ox.ac.uk}
\affil[4]{sean.bohun@ontariotechu.ca}

\begin{abstract}
	Simple mathematical models have been unable to predict the conditions leading to instabilities in a tuneable ring laser. Here, we propose a nonlinear iterative map model for tuneable ring lasers. Solving a reduced nonlinear Schr\"odinger equation for each component in the laser cavity, we obtain an algebraic map for each component. Iterating through the maps gives the total effect of one round trip. By neglecting the nonlinearity, we find a linearly chirped Gaussian to be the analytic fixed point solution, which we analyze asymptotically. We then numerically solve the full nonlinear model, allowing us to probe the underlying interplay of dispersion, modulation, and nonlinearity as the pulse evolves over hundreds of round trips of the cavity. In the nonlinear case, we find the chirp saturates, and the Fourier transform of the pulse becomes more rectangular in shape. Finally, for a nominal plane in the parameter space, we uncover a rich, sharp boundary separating the stable region and the unstable region where modulation instability degrades the pulse into an unsustainable state.
\end{abstract}

\setboolean{displaycopyright}{true}

\begin{document}

\maketitle

\section{Introduction}
\label{sec:intro}

\begin{figure}[tbp]
	\centering
	\begin{tikzpicture}
		\draw [rounded corners=4mm] (0,0) rectangle ++(6,4);
		\draw [rounded corners=4mm] (0,0) rectangle ++(-1.5,4);
	
		\draw (0.5,2.25) circle (0.5cm);
		\draw (0.5,2) circle (0.5cm) node [anchor=west,xshift=0.5cm,align=center] {Er-doped \\ gain fibre};
		\draw (0.5,1.75) circle (0.5cm);
	
		\filldraw[fill=white, draw=black] (2,-0.75) rectangle ++(2,1.5) node [midway] {Modulator};
		\filldraw[fill=white, draw=black] (-2.1,1.5) rectangle ++(1.2,1) node [midway] {Pump};
	
		\draw[-stealth] (3,4) -- (3,5.5) node [pos=0.75,anchor=west,xshift=0.25cm] {Laser output};
		\draw[densely dashdotted] (2.5,3.5) -- (3.5,4.5) node [pos=1,anchor=north,yshift=-0.75cm,align=center] {Optical \\ coupler};
	
		\filldraw[fill=white, draw=black] (6,2) circle (0.5cm);
		\draw[->,>=stealth] (6,2.325) arc (90:360:0.325cm);
	
		\filldraw[fill=white, draw=black] (0,0) circle (0.5cm);
		\draw[->,>=stealth] (0,0.325) arc (90:360:0.325cm);
	
		\filldraw[fill=white, draw=black] (0,4) circle (0.5cm) node [anchor=south,align=center,yshift=0.5cm] {Optical circulator};
		\draw[->,>=stealth] (0,4.325) arc (90:360:0.325cm);
	
		\draw [->,>=stealth,domain=20:70,blue] plot ({0.675*cos(\x)}, {0.675*sin(\x)});
		\draw [->,>=stealth,domain=110:160,red] plot ({0.675*cos(\x)}, {0.675*sin(\x)});
		\draw [->,>=stealth,domain=110:160,blue] plot ({6+0.675*cos(\x)}, {2+0.675*sin(\x)});
		\draw [->,>=stealth,domain=200:250,blue] plot ({6+0.675*cos(\x)}, {2+0.675*sin(\x)});
		\draw [->,>=stealth,domain=200:250,red] plot ({0.675*cos(\x)}, {4+0.675*sin(\x)});
		\draw [->,>=stealth,domain=290:340,blue] plot ({0.675*cos(\x)}, {4+0.675*sin(\x)});
	
		\draw [->,>=stealth,domain=270:360,blue] plot ({2.325+0.5*cos(\x)}, {4.675+0.5*sin(\x)});
		\draw [->,>=stealth,blue] (2.125, 4.675-0.5) -- (2.325+0.6, 4.675-0.5);
	
		\draw (5.5,2) -- (3.5,2) node [pos=0.5,anchor=south,yshift=0.25cm,xshift=-0.15cm] {CFBG};
		\foreach \i in {0,...,13}
			\draw (3.5 + \i*\i/100,1.75) -- (3.5 + \i*\i/100,2.25);
	\end{tikzpicture}
	\caption{Common cavity construction of a fibre ring laser~\cite{burgoyne2014, chung2017, lapre2019, shao2019}. The pump laser travels clockwise around the left-hand loop (red arrows) energizing the Er-doped gain fibre. The main laser travels clockwise around the right-hand loop (blue arrows), with most of the pulse exiting the cavity through the optical coupler.}
	\label{fig:cavity}
\end{figure}
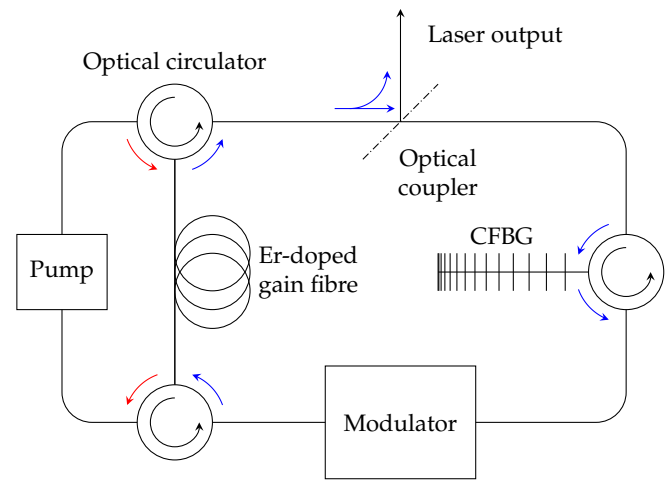

The sophistication of current active research for dispersion-tuned actively mode-locked lasers lies in stark contrast to current modelling efforts. These lasers can have ultrashort pulses, as short as 50 femtoseconds \cite{chung2017}, and the ability to tune their output frequency rapidly over a range of over 100 nm~\cite{bohun2015, burgoyne2010, chung2017, yamashita2009}. The capability to alter the operating frequency makes tuneable lasers particularly useful for imaging applications, such as optical coherence tomography~\cite{bohun2015, burgoyne2014, yamashita2009}, coherent anti-Stokes Raman spectroscopy~\cite{burgoyne2014}, and deep tissue multi-photon microscopy~\cite{chung2017}, as well as sensing and measuring of other ultrafast processes~\cite{burgoyne2014, silfvast2004, oktem2010}. Tuneable lasers are usually constructed in a ring with five main components: an optical coupler, a chirped fibre Bragg grating (CFBG), a modulator (or saturable absorber), a gain fibre (usually doped with Er), and a pump laser. A typical ring laser cavity is depicted in Fig.~\ref{fig:cavity}. The Kerr nonlinearity plays a crucial role in the dynamics within the cavity due to the high power and ultrashort duration of the~pulses.

Several effects arise within the cavity due to the interplay of dispersion, modulation, and nonlinearity~\cite{bohun2015, coen1997, lapre2019, meng2020, oktem2010, shao2019, woodward2018}. The two effects of most interest for us are wave breaking, and modulation instability. Wave breaking causes the leading edge of a pulse to redshift, and the trailing edge to blueshift (the effect reverses in the anomalous dispersion regime) to the point a shock develops~\cite{anderson1992, rothenberg1989a, rothenberg1989b, tomlinson1984, tomlinson1985}. The frequency shift causes the pulse to become more rectangular in the frequency domain, with a linear chirp over most of the pulse, saturating at the leading and trailing edge. In optics, wave breaking arises from self-phase modulation (SPM), which is a direct effect of the Kerr nonlinearity, causing the pulse to interfere with itself. New frequencies are then generated such that a rectangular profile is induced~\cite{agrawal2013, woodward2018}. Modulation instability---which typically arises in the anomalous dispersion regime---is the other effect that will be of interest. However, in the presence of multiple frequencies, modulation instability can emerge in the normal dispersion regime as well through cross-phase modulation and four-wave mixing~\cite{agrawal1987, agrawal2013, haelterman1992}. Modulation instability causes a pulse to degrade into multiple, much shorter, less intense pulses and can cause a laser to lose coherence~\cite{agrawal1987, coen1997, haelterman1992}. Outside of optics, wave breaking and modulation instability can occur in other areas with nonlinear waves and dispersive media such as plasma physics, heat propagation, transmission lines, and fluid dynamics~\cite{coen1997, rothenberg1989b}. Within a ring laser cavity, wave breaking and modulation instability can become parasitic, leading to an unstable and unsustainable pulse. Understanding the rich landscape of the parameter space is important for determining design principles to guarantee the ring laser is stable and sustainable~\cite{bohun2015, burgoyneemail, finot2008, lapre2019, woodward2018}.

The canonical equation in nonlinear optics is the nonlinear Schr\"odinger equation,
\begin{align}
	\pdiff{A}{z} &= - i \frac{\beta_2}{2}\pdiff[2]{A}{T} + i \gamma |A|^2 A.
	\label{eq:smallnlse}
\end{align}
Here, $A = A(T, z)$ is the complex pulse envelope, $T$ is the co-moving time, and $z$ is the propagation coordinate. The respective parameters, $\beta_2$, and $\gamma$, are the group velocity dispersion and the Kerr nonlinearity. Typically, \eqref{eq:smallnlse} is generalized to incorporate terms more specific to laser physics. A gain term and a loss term are added (and sometimes higher order dispersion terms), to yield the generalized nonlinear Schr\"odinger equation (GNLSE)~\cite{agrawal2013, bohun2015, finot2008, peng2018, shtyrina2017, yarutkina2013},
	\begin{align}
	\pdiff{A}{z} &= - i \frac{\beta_2}{2}\pdiff[2]{A}{T} + i \gamma |A|^2 A + \frac{1}{2}g(A) A - \alpha A,
	\label{eq:nlse}
\end{align}
where $g(A)$ is an amplifying term owing to the gain fibre, and $\alpha$ is the loss due to scattering and absorption.

The GNLSE, \eqref{eq:nlse}, has been applied in nonlinear optics~\cite{agrawal2013} and fibre optic communications~\cite{agrawal2002}. However, \eqref{eq:nlse} is regularly generalized even further to include a modulation term. The resulting equation is referred to as the master equation of mode-locking (see \cite{haus1975, haus1984, haus2000, tamura1996, usechak2005}), or sometimes as the Ginzburg--Landau equation. Owing to the nonlinear term in \eqref{eq:nlse} and the often non-trivial form of $g(A)$, as well as the possible addition of a modulation term, no analytic solution is known to the master equation or to \eqref{eq:nlse}. A comprehensive description of the theory and history of mode-locking lasers is presented in Haus~\cite{haus2000}. While the GNLSE well represents the dynamics in a single optical fibre, \eqref{eq:nlse} is not entirely representative of the underlying physics within a laser cavity such as Fig. \ref{fig:cavity}. The derivation of \eqref{eq:nlse} assumes that each process affects the pulse continuously within the cavity. The assumption that a single GNLSE applies to the whole laser cavity breaks down if the pulse undergoes large changes during any one round trip. Often, this assumption does not hold because of the specialized, discrete components.

To resolve some of the issues with an `average' model, each component and region in the laser is modelled with a separate GNLSE. The solution of one GNLSE is used as the initial condition for the GNLSE for the next component~\cite{lapre2019, meng2020, oktem2010, woodward2018}. An iterative map model allows for better encapsulation of specific dynamics introduced by each component. The main drawback of this method is the extreme computational expense since a nonlinear partial differential equation must be numerically solved for every segment for each of the hundreds of round trips in a given simulation. We can alleviate this computational expense by noticing each effect is localized to a corresponding component in the cavity. A majority of the loss is due to the optical coupler, and virtually all the dispersion happens within the CFBG~\cite{agrawal2002}. The pulse is only modulated by the modulator, and the pulse is amplified only within the Er-doped gain fibre. Therefore, we can drastically simplify each of the GNLSEs into a form we can solve analytically. Using analytic maps instead of numerically solving a GNLSE reduces the computational cost, while maintaining the benefit of reflecting the specific geometry of a component-wise~simulation.

Iterative map methods have been used for optics since 1955 when Cutler~\cite{cutler1955} analyzed a microwave regenerative pulse generator. Siegman and Kuizenga~\cite{kuizenga1970a, kuizenga1970b, kuizenga1970, siegman1969} adapted Cutler's method for both AM and FM mode-locked lasers in 1969, and Martinez \emph{et al.}~\cite{martinez1984, martinez1985} incorporated the nonlinearity while modelling passively mode-locked lasers. A linear iterative method was used for dispersion-tuned actively mode-locked lasers recently by Burgoyne \emph{et al.}~\cite{burgoyne2014}. Each of these models used a prescribed transfer function to describe the effect each component has on the pulse. While the transfer functions are chosen suitably to have the effect observed in the laboratory, the transfer functions are phenomenologically chosen, and not necessarily obtained directly from the underlying physics. Furthermore, these methods have not included the effect of every component, lacking the interactions of dispersion, modulation, and nonlinearity---hindering their models' ability to exhibit modulation instability.

In this paper, we build an iterative map model for the evolution of a laser pulse as the pulse travels through the cavity. In Section~\ref{sec:model}, we derive our own iterative map from \eqref{eq:nlse}, we non-dimensionalize the model and discover four non-dimensional parameters that govern the dynamics of the system. We then examine the results of our model in Section~\ref{sec:results}---first solving the model analytically by assuming the absence of the nonlinearity, then numerically solving the full nonlinear model. A sharp boundary is identified in the parameter space separating the region of stability, and the region where modulation instability degrades the pulse. Finally, our concluding thoughts and ideas for possible future work are given in Section~\ref{sec:conclusion}.

\section{Iterative Map Model}
\label{sec:model}
We derive our own iterative map model for the evolution of a pulse through a laser cavity by solving a reduced GNLSE for each component in the cavity---except the modulator, which we consider to be applied externally by an electro-optic modulator. Starting with the gain component, we reduce \eqref{eq:nlse} to 
\begin{align}
	\pdiff{A}{z} &= \frac{1}{2}g(A)A,
	\label{eq:gainexample}
\end{align}
by assuming all other effects are negligible. Once again, our assumption is justified by the localization of each effect to its corresponding component, allowing us to simplify \eqref{eq:nlse} to \eqref{eq:gainexample} in the gain fibre. To proceed, we assume the gain has the form
\begin{align}
	g(A) &= \frac{g_0}{1 + E / \Es},
	\label{eq:gainform}
\end{align}
where $g_0$ is a small signal gain, $E$ is the energy of the pulse, defined as
\begin{align}
	E &= \int_{-\infty}^\infty |A|^2 \, \df T,
	\label{eq:energy}
\end{align}
and $\Es$ is the energy that the gain begins to saturate~\cite{haus1984, shtyrina2017, silfvast2004}, which depends on the power of the pump laser. With the choice of $g(A)$ given in \eqref{eq:gainform}, \eqref{eq:gainexample} becomes
\begin{align}
	\pdiff{A}{z} &= \frac{g_0 A}{2 \left( 1 + E / \Es \right)}.
	\label{eq:pdiffgain}
\end{align}
Multiplying \eqref{eq:pdiffgain} by the complex conjugate of $A$, we have
\begin{align}
	\overline{A} \pdiff{A}{z} &= \frac{g_0}{2 \left( 1 + E / \Es \right)} |A|^2.
	\label{eq:pdiffgain2}
\end{align}
By adding \eqref{eq:pdiffgain2} and the complex conjugate of \eqref{eq:pdiffgain2}, and then integrating over $T$, we have
\begin{align}
	\diff{E}{z} &= \frac{g_0 E}{1 + E / \Es}.
	\label{eq:energyde}
\end{align}
By assuming the energy entering the gain fibre is $E_g$, and the energy after travelling through the gain fibre of length $L_g$ is $E_{\text{out}}$, we integrate \eqref{eq:energyde} to find the energy after amplification is
\begin{align}
	\label{eq:engain}
	E_{\text{out}} &= \Es W \left( \frac{E_g}{\Es} \exp \left( \frac{E_g}{\Es} \right) \exp( g_0 L_g ) \right),
\end{align}
where $W$ is the Lambert $W$ function. Moreover, as the gain will amplify the pulse without changing the shape of the envelope, the pulse after travelling through the gain fibre can be expressed as
\begin{align}
	G(A) &= \mathcal{G} A.
	\label{eq:ampfact}
\end{align}
By \eqref{eq:energy}, \eqref{eq:engain} and \eqref{eq:ampfact},
\begin{align}
	\mathcal{G}^2 E_g &= E_\text{out},
\end{align}
therefore,
\begin{align}
	G(A) &= \left( \frac{E_\text{out}}{E_g} \right)^{1/2} A \\
	&=	\left( \frac{\Es}{E_g} W \left( \frac{E_g}{\Es} \exp \left( \frac{E_g}{\Es} \right) \exp( g_0 L_g ) \right) \right)^{1/2} A \label{eq:dimgain}
\end{align}
for the amplification of the pulse after the Er-doped gain fibre.

Repeating the same procedure for the nonlinearity, loss, and dispersion components yields
\begin{align}
	\pdiff{A}{z} &= i \gamma |A|^2 A, \label{eq:nlde} \\
	\pdiff{A}{z} &= - \alpha A,  \label{eq:lossde} \\
	\pdiff{A}{z} &= - i \frac{\beta_2}{2} \pdiff[2]{A}{T} \label{eq:dispde}
\end{align}
as the reduced versions of \eqref{eq:nlse}, respectively. Solving \eqref{eq:nlde} over a length of fibre $L_f$, we find the effect of the fibre nonlinearity is
\begin{align}
	F(A) &= A \exp \left( i \gamma |A|^2 L_f \right). \label{eq:dimnl}
\end{align}
Similarly, the effect of the loss is
\begin{align}
	L(A) &= (1 - R) \exp( - \alpha L_T )A, \label{eq:dimloss}
\end{align}
where $L_T$ is the total length of fibre in the cavity, $R$ is the reflectivity of the optical coupler, and the $(1 - R)$ term  is the loss to the output of the laser. The effect of dispersion is obtained by solving \eqref{eq:dispde} over the `length' of the dispersive medium, $L_D$,
\begin{align}
	D(A) &= \FTi{\exp( i \omega^2 \beta_2 L_D / 2 ) \FT{A}}, \label{eq:dimdisp}
\end{align}
where, $\mathcal{F}$ and $\mathcal{F}^{-1}$ denote the Fourier transform and its inverse.

Finally, we consider the modulation. We assume the pulse is modulated by an electro-optic modulator. For simplicity, the form is taken as the Gaussian
\begin{align}
	M(A) &= \exp( -T^2 / 2 T_M^2 ) A,
	\label{eq:modform}
\end{align}
where $T_M$ is the characteristic width of modulation. This assumption is not restrictive. The results apply to any square integrable modulation function through a result of Calcaterra and Boldt~\cite{calcaterra2008a} who prove that linear combinations of translations of a single Gaussian, $\textrm{e}^{-x^2}$, are dense in $L^2(\mathbb{R})$. Any continuous modulation function with compact support would fall into this class. Additionally, using a saturable absorber in place of an electro-optic modulator is common in practice. In this case, we can exchange \eqref{eq:modform} for~\cite{lapre2019, meng2020, oktem2010, woodward2018}
\begin{align}
	S(A) &= \left( 1 - \frac{q_0}{1 + |A|^2 / P_\text{sat}} \right) A,
\end{align}
where $q_0$ is the modulation depth, and $P_\text{sat}$ is the saturation power.

\subsection{Combining the Component Maps}
\label{sec:effects}
Now that we have the algebraic effect of each component of the cavity, we are ready to take the composition of the maps to give the effect of one round trip. Unlike the models described in Section \ref{sec:intro}, the order of our components is important since the maps in \eqref{eq:dimgain}, and \eqref{eq:dimnl}--\eqref{eq:modform} do not necessarily commute with each other.

\begin{figure}[tbp]
	\centering
	\includegraphics{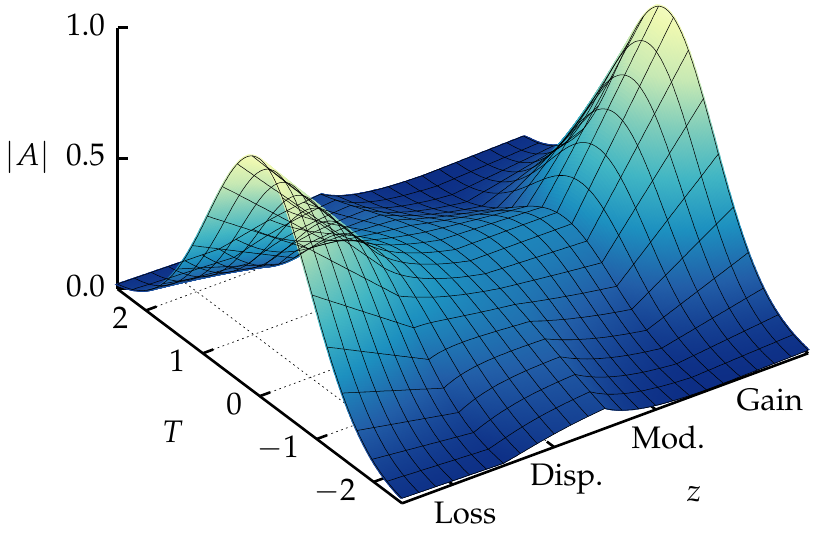}
	\caption{Evolution of the envelope during one round trip of the cavity. The pulse decays because of the optical coupler, disperses because of the CFBG, is modulated by the modulator, and finally, is amplified by the gain fibre (the envelope is unaltered by the nonlinearity). Note, the $z$ axis is not to scale.}
	\label{fig:cavityevo}
\end{figure}

The geometry of a particular laser will dictate the order of the process maps. However, we assume the construction of the laser is as shown in Fig.~\ref{fig:cavity}. The effect of the nonlinearity will be maximal after the pulse passes through the gain fibre---the pulse is most energetic after being amplified by the gain. Thus, we take the nonlinearity component to follow the gain. Similarly, we want the loss component to directly follow the nonlinearity in an attempt to minimize the effect the nonlinearity has on the pulse. All that remains is the dispersion and modulation components. For now, let us assume the CFBG precedes the modulator. Finally, we assume the loss component is our starting point since the loss corresponds to the optical coupler, and thus, allows us to more easily compare our results with experiments. More precisely, let
\begin{align}
	\mathcal{L}(A) = F(G(M(D(L(A)))))
	\label{eq:order}
\end{align}
denote one complete round trip of the cavity. Since a ring laser is cyclic in nature, we use the result of one round trip as the input pulse for the next round trip---restarting the process. We continually repeat this procedure until the envelope and chirp of the pulse reach equilibrium, or the pulse succumbs to modulation instability. The phase change is uninteresting to us, as the phase cannot easily be measured experimentally. An illustration of the evolution of the envelope during one round trip of the laser cavity can be found in Fig.~\ref{fig:cavityevo}.

\begin{table*}[!b]
	\centering
	\caption{Range of parameter values (taken from~\cite{agrawal2013, burgoyne2014, burgoyneemail, tamura1996, usechak2005}).}
 	\label{tab:values}
 	\begin{tabular}{clll}
		\hline
		Parameter & Description & Value & Units \\
		\hline\noalign{\smallskip}
		$T_M$ & Characteristic modulation time & $15$--$150$ & ps \\
		$\beta_2 L_D$ & Dispersion provided by the CFBG & $10$--$2000$ & ps$^2$ \\
		$\gamma$ & Kerr nonlinearity coefficient & $0.001$--$0.01$ & W$^{-1}$ m$^{-1}$ \\
		$L_f$ & Length of fibre between gain and optical coupler & $0.15$--$1$ & m \\
		$L_g$ & Length of the Er-doped gain fibre & $2$--$3$ & m \\
		$\alpha$ & Loss of fibre due to scattering and absorption & $10^{-4}$--$0.3$ & m$^{-1}$ \\
		$R$ & Reflectivity of optical coupler & $0.1$--$0.9$ & -- \\
		$\Es$ & Saturation energy of Er-doped gain fibre & $10^3$--$10^4$ & pJ \\
		$g_0$ & Small signal gain &  $1$--$10$ & m$^{-1}$ \\
		$L_T$ & Total length of fibre in the cavity & $10$--$100$ & m \\
		\noalign{\smallskip}\hline
	\end{tabular}
\end{table*}

\subsection{Non-Dimensionalization}
Non-dimensionalization is a technique used in mathematical modelling to better understand the relative importance of different processes within a system. In essence, we scale all variables by typical values based on the geometry and boundary conditions (see, for example,~\cite{howison2005}). Doing so `factors' the dimensional units out of the problem---leaving only dimensionless equations with dimensionless parameters. The relative magnitudes of the dimensionless parameters characterize the dominant dynamics---comparable to how the Reynolds number describes fluid flow. Each process in the laser can be better understood by scaling the time, energy, and amplitude by the convenient factors:
\begin{align}
	T &= T_M \widetilde{T},& E &= \Es \widetilde{E},& A &= \left( \frac{\Es}{T_M} \right)^{1/2} \widetilde{A}.
\end{align}

We find the non-dimensional mappings (after dropping the tildes) are
\begin{equation}
	\begin{aligned}
		G(A) &= \left(E_g^{-1} W \left( a E_g \textrm{e}^{E_g}\right) \right)^{1/2} A,\\
		F(A) &= \exp \left( i b |A|^2 \right) A, \\
		L(A) &= h A, \\
		D(A) &= \FTi{\exp \left( i s^2 \omega^2 \right) \FT{A}}, \\
		M(A) &= \exp \left( -T^2 / 2 \right) A.
	\end{aligned}
	\label{eq:effects}
\end{equation}
where we have introduced four dimensionless parameters, namely
\begin{equation}
	\hspace*{-1mm}
	\begin{aligned}
		a &= \exp( g_0 L_g ) \sim 8 \times 10^3, & 
		b &= \gamma L_f \frac{\Es}{T_M} \sim 1, \\
		h &= (1 - R) \exp( - \alpha L_T) \sim 0.04, &
		s &= \left( \frac{\beta_2 L_D}{2 T_M^2} \right)^{1/2} \sim 0.2,	
	\end{aligned}
	\label{eq:ndparam}
\end{equation}
and we have estimated their sizes using values in Table~\ref{tab:values}.

Each expression in \eqref{eq:effects} contains a single parameter that characterizes the device. The parameter governing the modulation, the time $T_M$, does not appear in $M(A)$, but is used to define the relative strength of the other devices. The gain is characterized by $a$, a measure of the energy added by the gain fibre; the Kerr nonlinearity by $b$, prescribing the frequency shift; the total loss by $h$, includes absorption and the optical coupler; and the dispersion by $s$, specifying the effectiveness of the CFBG to the width of the pulse.

\section{Results}
\label{sec:results}
We split the results into two subsections. First, we investigate the small nonlinearity limit ($b \ll 1$), and then we explore the full nonlinear model.

\subsection{Linear Solution}
We neglect the effect of the nonlinearity, that is, we take $b \ll 1$. Proceeding by assuming a fixed point solution in the representation of a linearly chirped Gaussian, our aim is to find the variance, chirp, and energy of the pulse. The reasons for this assumption are that the solution to the models presented in~\cite{cutler1955, siegman1969, kuizenga1970a, martinez1984, martinez1985} were Gaussian, the equilibrium shape will correlate to the modulation function, and a Gaussian is a fixed point of the Fourier transform. Furthermore, we choose a linearly chirped Gaussian because it resembles the envelope, and linear chirp expected from the literature~\cite{burgoyne2014, haus1975, haus1996, haus2000, usechak2005}.

\begin{figure}[tbp]
	\centering
	\includegraphics{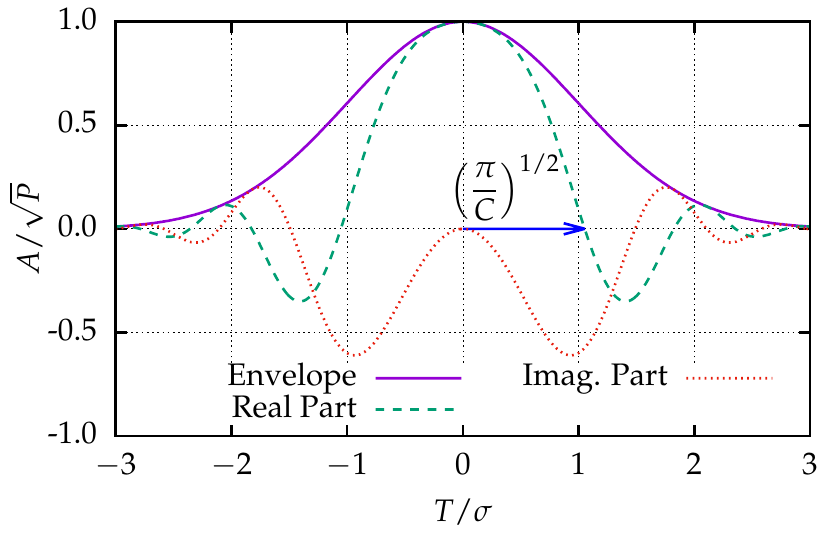}
	\caption{Graph of the linearly chirped Gaussian pulse given by \eqref{eq:A0} with $\phi_0 = 0$.}
	\label{fig:samplegauss}
\end{figure}

\begin{figure*}[tbp]
	\centering
	\begin{subfigure}{\columnwidth}
		\centering
		\includegraphics{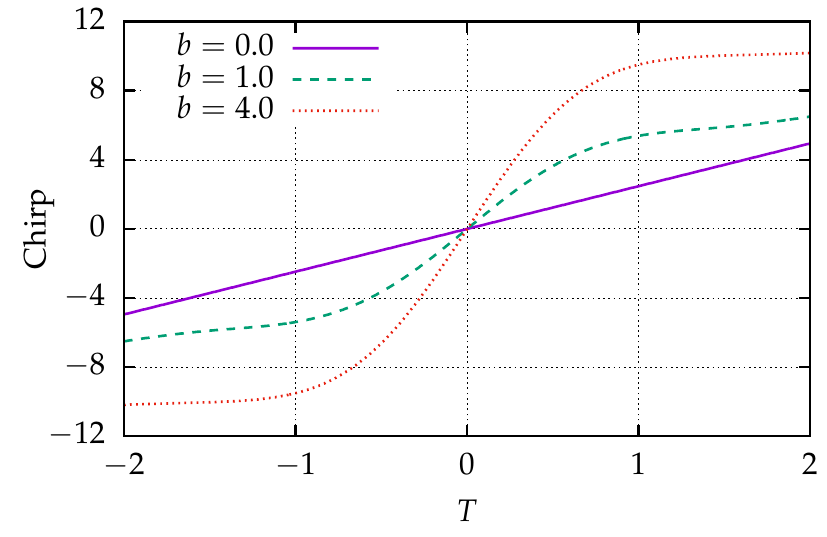}
		\caption{Chirp.}
		\label{fig:chirp}
	\end{subfigure} %
	\begin{subfigure}{\columnwidth}
		\centering
		\includegraphics{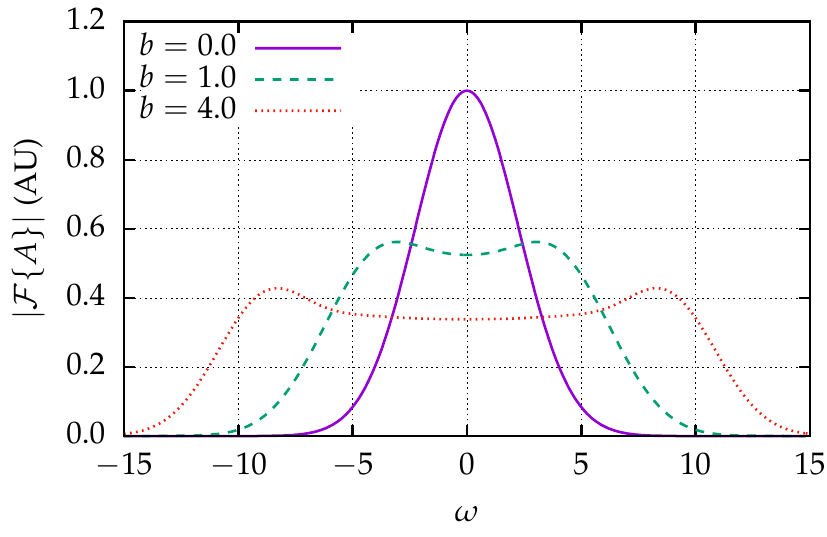}
		\caption{Fourier transform.}
		\label{fig:ft}
	\end{subfigure}
	\caption{Chirp and Fourier transform at equilibrium (after 500 round trips) initiated by \eqref{eq:nlA0} for three values of $b$, the nonlinearity parameter ($a = 8 \times 10^3$, $h = 0.04$, $s = 0.2$, and $E_0 = 0.1$).}
	\label{fig:chirpft}
\end{figure*}

We consider the pulse, shown in Figure \ref{fig:samplegauss}, given by
\begin{align}
	A = \sqrt{P} \exp \left( -(1 + iC) \frac{T^2}{2 \sigma^2} \right) \exp(i \phi_0),
	\label{eq:A0}
\end{align}
where $P$ is the peak power, $C$ is the chirp, $\sigma^2$ is the variance, and $\phi_0$ is the phase. By computing \eqref{eq:A0} after one round trip, and solving $\mathcal{L}(A) = A \textrm{e}^{i \phi}$, tedious manipulations yield the conditions
\begin{align}
	\frac{\sigma^4}{1 - \sigma^2} &= \left( \sigma^2 + 2 C s^2 \right)^2 + 4 s^4, \label{eq:sigrelation} \\
	\frac{C}{1 - \sigma^2} &= C + 2 \frac{s^2}{\sigma^2} \left( 1 + C^2 \right), \label{eq:chirprelation} \\
	1 &= \frac{W(a E_g \textrm{e}^{E_g})}{E_g} h^2 \left( 1 - \sigma^2 \right)^{1/2} \label{eq:energyrelation}
\end{align}
for \eqref{eq:A0} to be in equilibrium. Using \eqref{eq:sigrelation} and \eqref{eq:chirprelation} we eliminate the chirp, $C$, and find an expression relating $\sigma^2$ and $s$,~namely
\begin{align}
	\sigma^8 + 4 s^4 \sigma^6 - 20 s^4 \sigma^4 + 32 s^4 \sigma^2 - 16 s^4 = 0.
	\label{eq:var}
\end{align}
\eqref{eq:var} has the positive, real, analytic solution
\begin{equation}
	\begin{split}
		\sigma^2 = \sqrt{2} s \left( s^6 + 3s^2 + \sqrt{4 + s^4} \left( 1 + s^4 \right) \right)^{1/2} & \\
		- s^4 - s^2 & \sqrt{4 + s^4}.
		\label{eq:equilvar}
	\end{split}
\end{equation}
We find the chirp
\begin{align}
	C = \frac{\sigma^4}{2 s^2 (2 - \sigma^2)}
	\label{eq:chirp}
\end{align}
by combining \eqref{eq:sigrelation} and \eqref{eq:chirprelation}. We obtain the energy of the pulse entering the gain fibre, $E_g$, from \eqref{eq:energyrelation}:
\begin{align}
	E_g &= \frac{h^2 \left( 1 - \sigma^2 \right)^{1/2} \log \left( a h^2 \left( 1 - \sigma^2 \right)^{1/2} \right)}{h^2 \left( 1 - \sigma^2 \right)^{1/2} - 1},
\end{align}
and thus, the equilibrium energy at the optical coupler, is
\begin{align}
	E &= W \left( a E_g \textrm{e}^{E_g} \right) \\
	&= \frac{\log \left( a h^2 \left( 1 - \sigma^2 \right)^{1/2} \right)}{1 - h^2 \left( 1 - \sigma^2 \right)^{1/2}}.
	\label{eq:analenergy}
\end{align}
The argument of the logarithm in \eqref{eq:analenergy} gives us the condition
\begin{align}
	a h^2 (1 - \sigma^2)^{1/2} > 1
	\label{eq:energycond}
\end{align}
for the pulse to be sustainable.

Equating the energy of the pulse given by \eqref{eq:A0} and \eqref{eq:analenergy}, we find the peak power to be
\begin{align}
	P &= \frac{\log \left( a h^2 \left( 1 - \sigma^2 \right)^{1/2} \right)}{\sqrt{\pi} \sigma \left( 1 - h^2 \left( 1 - \sigma^2 \right)^{1/2} \right)}.
	\label{eq:analpower}
\end{align}

The forms of \eqref{eq:equilvar}--\eqref{eq:analpower} make the expressions difficult to work with. However, recall that $s$ is small, thus, we asymptotically expand \eqref{eq:equilvar}--\eqref{eq:analpower} assuming $s \ll 1$, and
\begin{align}
	\sigma^2 &\sim 2s(1 - s) + \bigO{s^3}, \\
	C &\sim 1 - s + \frac{1}{2}s^2 + \bigO{s^3}, \\
	E &\sim \frac{\log (a h^2)}{1 - h^2} - \frac{1}{1 - h^2} \left( 1 + \frac{h^2 \log(a h^2)}{1 - h^2}  \right) s + \bigO{s^2}, \label{eq:maxenergy} \\
	P &\sim \frac{\log(ah^2)}{\sqrt{2 \pi}(1 - h^2)} s^{-1/2} + \bigO{s^{1/2}},
\end{align}
in the limit $s \rightarrow 0$. The expansions in the limit $s \rightarrow \infty$ are given in Appendix \ref{sec:largedisp}.

Our linear model exhibits a solution of the same form as previous linear models. However, by including the effect of the nonlinearity, we uncover the rich interplay between dispersion, modulation, and nonlinearity.

\subsection{Nonlinear Solution and Instability}
\label{sec:nlresults}
The nonlinear solution is solved with a numerical simulation. For the input pulse, one of the common forms is a hyperbolic secant~\cite{coen1997, finot2008, rothenberg1989b, tomlinson1984}, which we assume has the exact form
\begin{align}
	A_0 = \Gamma \sech \left( 2 T \right) \textrm{e}^{i \pi / 4}.
	\label{eq:nlA0}
\end{align}
Here, $\Gamma$ is a normalizing factor chosen so the pulse has the initial energy $E_0$, which we take as $E_0 = 0.1$. Using \eqref{eq:nlA0} as a seed, we simulate the pulse travelling around the cavity using $2^{18}$ points, and a time span of 16.

In Fig.~\ref{fig:chirpft} we show the chirp and Fourier transform of the pulse \eqref{eq:nlA0} after reaching equilibrium for the nominal parameter values given in \eqref{eq:ndparam}, and two additional $b$ values. In the nonlinear case, we find the equilibrium pulse is no longer Gaussian---as evidenced by the Fourier transforms. Notice, in comparison to the linear case, the nonlinearity introduces higher frequency modes, giving a bi-modal distribution. Moreover, we find the pulse is linearly chirped near the peak, but, in the tails, $|T| \gtrsim 1$, the chirp begins to saturate---consistent with the experimental results~\cite{chen2008, rothenberg1989b, tomlinson1985}. As the nonlinearity parameter, $b$, increases, the chirp increases more sharply across the pulse, and saturates at a larger value. Also, the two peaks in the Fourier transform become more pronounced and further apart, and the Fourier transform of the pulse becomes more rectangular. However, after a critical point, SPM plays a more substantial role, leading to modulation instability.

\begin{figure}[tbp]
	\centering
	\includegraphics{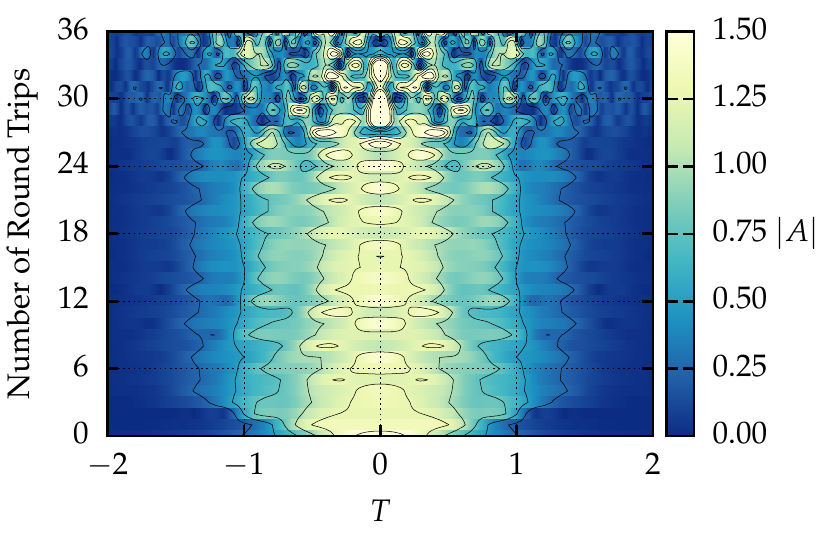}
	\caption{Example of a pulse destabilizing from the initial pulse \eqref{eq:nlA0} ($a = 8 \times 10^3$, $h = 0.04$, $b = 1.6$, $s = 0.1$, and $E_0 = 0.1$).}
	\label{fig:breakevo}
\end{figure}

\begin{figure}[tbp]
	\centering
	\includegraphics{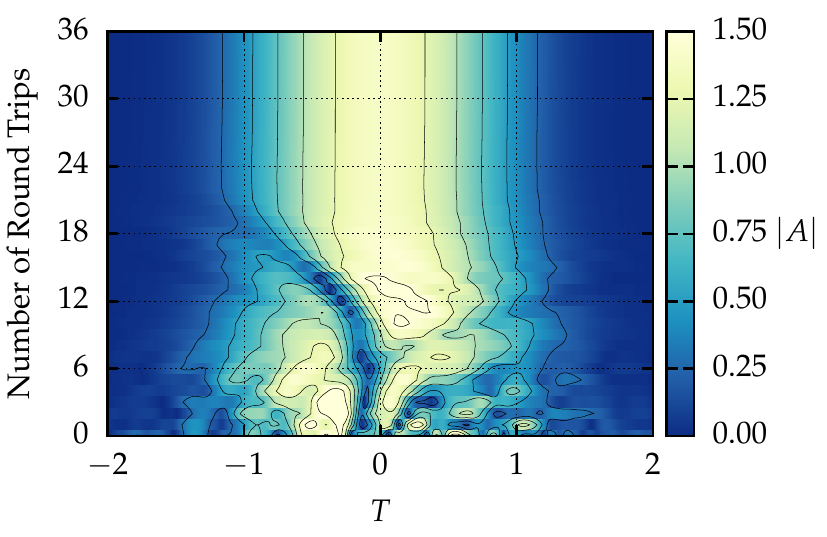}
	\caption{Example of a pulse coming to equilibrium from noise ($a = 8 \times 10^3$, $h = 0.04$, $b = 1.0$, $s = 0.1$, and $E_0 = 0.1$).}
	\label{fig:convevo}
\end{figure}

Fig.~\ref{fig:breakevo} highlights the phenomenon of modulation instability. By decreasing the dispersion compared to Fig.~\ref{fig:chirpft}, we find the pulse begins to `breathe'. During the first two dozen round trips of the cavity the SPM compounds and becomes detrimental to the pulse. In turn, modulation instability is induced, and degrades the pulse, until the pulse is no longer stable or sustainable. On the other hand, having a more moderate value of $b$, the pulse is able to equilibrate even with a less favourable, random, initial pulse (Fig.~\ref{fig:convevo}), as in~\cite{meng2020}. The pulse is able to shed the initial left lobe ($T \approx -1 / 2$) because of dispersion and modulation, and a central lobe ($T \approx 0$) forms, which is able to grow because of the gain medium. The central lobe then allows the pulse to come to equilibrium quickly. Furthermore, since the intensity of the initial pulse is small relative to the saturation energy, the effect of SPM is negligible during the first few round trips, allowing the laser to select preferable modes. The preferable modes then get amplified, stabilizing the pulse. Evidently, the initial shape of the pulse is much less important than the initial energy of the pulse, or the strength of the nonlinearity.

\begin{figure}[tbp]
	\centering
	\includegraphics{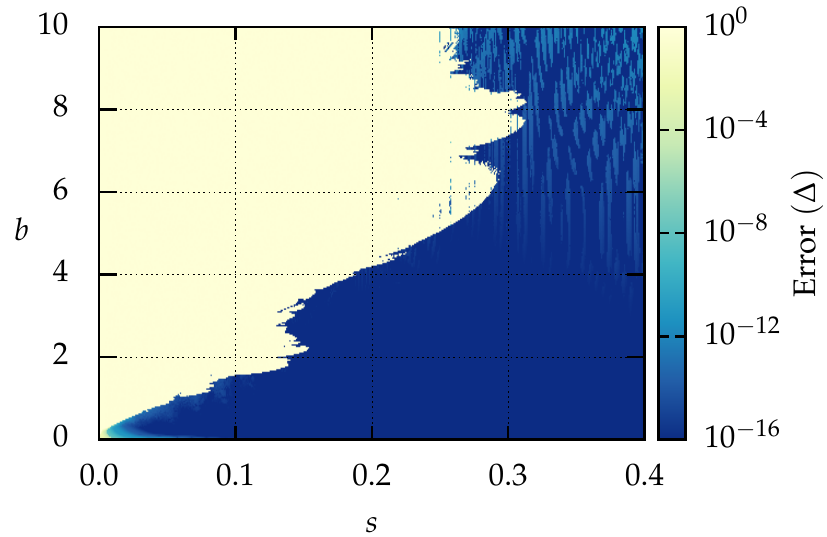}
	\caption{Relative error of a pulse's envelope between round trips 499 and 500 ($a = 8 \times 10^3$, $h = 0.04$, and $E_0 = 0.1$).}
	\label{fig:error}
\end{figure}

We now turn our attention to a more quantitative analysis, and characterize the stability of a laser by exploring the parameter space. We focus on the $s$--$b$ plane in particular, since $a$ and $h$ affect only the amplitude of the pulse, whereas the interactions of dispersion and nonlinearity give rise to the instabilities. We examine the relative change in the pulse's envelope between consecutive round trips of the cavity to identify stability. We compute the error as
\begin{align}
	\Delta = \frac{\| |\mathcal{L}^n(A_0)| - |\mathcal{L}^{n-1}(A_0)| \|_2}{\| \mathcal{L}^{n-1}(A_0) \|_2},
	\label{eq:error}
\end{align}
where
\begin{align}
	\| f \|_2^2 \coloneqq \int_{-\infty}^\infty |f|^2 \, \df T,
\end{align}
and we choose $n$ to be sufficiently large to guarantee the pulse reaches equilibrium or deteriorates because of modulation instability---we use $n = 500$ in our experiments. We take the modulus of the pulse since we are interested in the evolution of only the envelope because of the auto-correlation methods used for experimental measurements. We plot the error in Fig.~\ref{fig:error}. Two distinct regions split the parameter space, divided by an incredibly sharp, and complicated boundary. In the upper-left region, the error is $\bigO{1}$; here, the nonlinearity is too strong, and the SPM induces modulation instability. In the lower-right region we find the opposite behaviour---the pulse reaches an equilibrium state and is stable to within machine precision. The increase in dispersion, coupled with modulation, allows the laser to balance the nonlinear effects. 

\begin{figure}[tbp]
	\centering
	\includegraphics{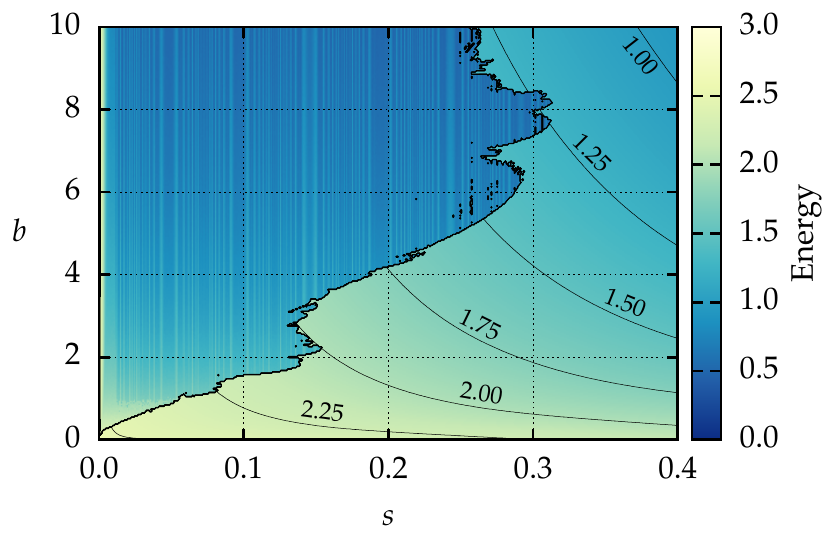}
	\caption{Energy of the pulse after 500 round trips ($a = 8 \times 10^3$, $h = 0.04$, and $E_0 = 0.1$).}
	\label{fig:energy}
\end{figure}

Also of interest is the energy of the pulse at equilibrium, which we compute numerically by \eqref{eq:energy}. Given our ordering of the components, we compute the energy of the pulse immediately before passing through the optical coupler. Thus, the energy of each pulse emitted by the laser is $(1 - h^2) E$, and $h^2 E$ remains in the cavity. We plot the energy in the $s$--$b$ plane in Fig.~\ref{fig:energy}. Unsurprisingly, we find the same sharp boundary as we saw in Fig.~\ref{fig:error}. In the unstable region (upper-left) the energy is small. The pulse is not able to foster prominent modes in this region, and is unsustainable, thus, the intensity is low across the entire pulse. In the stable region (lower-right) we find the energy decays smoothly as both $b$ and $s$ increase, with the contours being hyperbolic-esque. As $s$ increases, the tails of the pulse increase in length due to the dispersion, and more energy is lost due to modulation. As $b$ increases, our choice of modulation function helps suppress the additional frequencies generated by SPM, ensuring the pulse remains band-limited. We find a more energetic laser requires weaker dispersion, however, the nonlinearity must be correspondingly low, otherwise modulation instability will destroy the pulse. Moreover, the maximum energy, which occurs in the limit where both $s$ and $b$ go to zero, is 2.5487. With our choice of parameters, \eqref{eq:maxenergy} predicts an equilibrium energy of 2.5535 at the origin, confirming the usefulness of the approximation.

\begin{figure}[tbp]
	\centering
	\includegraphics{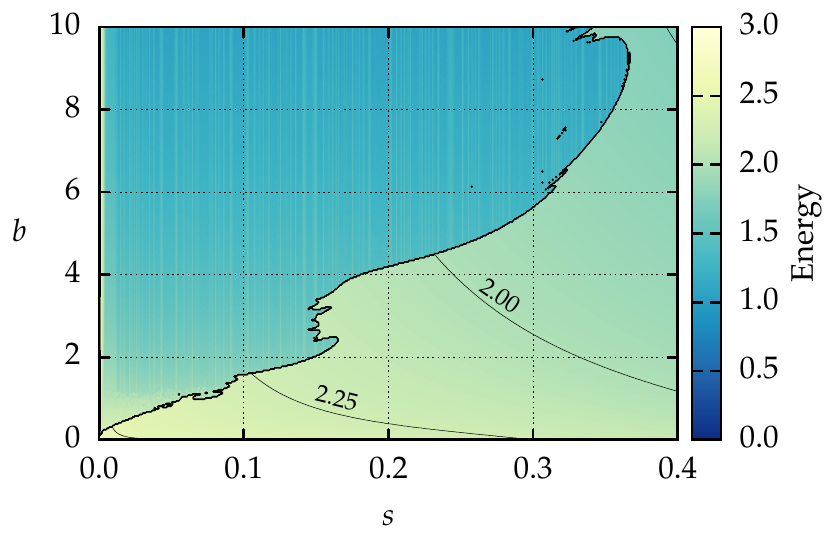}
	\caption{Energy of the pulse after 500 round trips with dispersion and modulation swapped ($a = 8 \times 10^3$, $h = 0.04$, and $E_0 = 0.1$).}
	\label{fig:energyswitch}
\end{figure}

Finally, we briefly investigate the affect the order of the components has on the dynamics. In Section~\ref{sec:model}\ref{sec:effects}, we justified the order of the gain, nonlinearity, and loss, but, just assumed dispersion preceded modulation. Now we consider the effect of swapping the dispersion and modulation components. The overall structure of the energy landscape (Fig.~\ref{fig:energyswitch}) is the same as in Fig.~\ref{fig:energy}. In the previous case, with modulation following dispersion, more energy is removed from the pulse tails. Whereas, having modulation first, more energy remains in the cavity. The additional energy makes the laser more powerful than in the previous case, which in turn increases the size of the unstable region. Comparing with Fig.~\ref{fig:energy}, the unstable upper-left region spreads farther to the right and upward, and the energy is larger in the stable region, especially away from the origin. Once again, the energy at the origin is 2.5487, as we are effectively removing the CFBG in the limit $s \rightarrow 0$.

\section{Conclusion}
\label{sec:conclusion}
Expanding upon the ideas originally proposed by Cutler~\cite{cutler1955}, and Kuizenga and Siegman~\cite{kuizenga1970, kuizenga1970a, siegman1969}, we developed a nonlinear iterative map model for tuneable ring lasers. We recovered linearly chirped Gaussian solutions in the case of a Gaussian modulation function when omitting the nonlinearity. Moreover, the solution of the linear model extends to any continuous function with compact support when using the results of~\cite{calcaterra2008a}. In contrast, with the inclusion of the nonlinearity, we were able to recover wave breaking and modulation instability, and found a sharp boundary of stability. These instabilities have been demonstrated in a laboratory setting~\cite{agrawal2013, anderson1992, finot2008, rothenberg1989b, tomlinson1985}, but, have proven difficult to predict with simple mathematical models~\cite{meng2020}. We have shown a simple iterative map model is able to reproduce complicated instabilities arising naturally from the interplay between dispersion, modulation, and nonlinear effects. Our model easily adapts to specific laser geometries, and to more complicated models of individual components. For example, incorporating a frequency dependence in the gain function, or model a saturable absorber instead of a Gaussian electro-optic modulator.

Analyzing the nonlinear model analytically using an asymptotic expansion for small values of the nonlinearity parameter, $b$, is of interest. Doing so will provide insight to how the nonlinearity impacts the linear solution to better understand the manifestation of wave breaking and modulation instability. Furthermore, a more detailed analysis of crossing the boundary between stability and instability may provide valuable insight for mechanisms that trigger the instability.

\appendix
\section{Large Dispersion Limit}
\label{sec:largedisp}
In the large dispersion limit, we have $s \rightarrow \infty$, and
\begin{align}
	\sigma^2 &\sim 1 - \frac{1}{4}s^{-4} + \bigO{s^{-8}}, \label{eq:siginfty} \\
	C &\sim \frac{1}{2}s^{-2} + \bigO{s^{-6}}.
\end{align}
We have no asymptotic expansion of the energy or peak power in the limit $s \rightarrow \infty$ since the pulse loses too much energy because of dispersion and modulation. The gain is not powerful enough to balance the energy lost. Expanding the requirement in \eqref{eq:energycond} using \eqref{eq:siginfty}, we find 
\begin{align}
	s^* = \left( \frac{a h^2}{2} \right)^{1/2}
	\label{eq:maxs}
\end{align}
is the maximum value of $s$ to facilitate a sustainable pulse.

\section*{Acknowledgement.}
Portions of this work were presented at The V AMMCS International Conference in 2019, ``A new method of modelling tuneable lasers with functional composition'' \cite{metherallammcs}. BM acknowledges the support provided by the EPSRC Centre for Doctoral Training in Industrially Focused Mathematical Modelling (EP/L015803/1). The authors would like to thank C. Breward for critical reading of the manuscript.

\section*{Disclosures.}
The authors declare no conflicts of interest.

\bibliography{Ref}

\end{document}